\documentclass[lettersize,journal]{IEEEtran}
\usepackage[utf8]{inputenc}
\usepackage[dvipsnames]{xcolor}
\usepackage{amsmath,amsfonts}
\usepackage{amssymb}
\usepackage{algorithmic}
\usepackage{array}
\usepackage[caption=false,font=normalsize,labelfont=sf,textfont=sf]{subfig}
\usepackage{subcaption}
\usepackage{caption}
\usepackage[%
    colorlinks=true,
    pdfborder={0 0 0},
    linkcolor=red
]{hyperref}
\usepackage{textcomp}
\usepackage{stfloats}
\usepackage{url}
\usepackage{verbatim}
\usepackage{graphicx}

\title{Feasibility of Negative Triangularity Equilibria in the SPARC Tokamak}
\author{Narin Y\"{u}ksek, Theodore Golfinopoulos}

\begin{document}
\maketitle
\begin{abstract}
We investigate the feasibility of negative triangularity (NT) plasma configurations in the SPARC tokamak, a compact, high-field device (up to 12.2~T) designed for positive triangularity (PT) operation. Using the FreeGS free-boundary equilibrium solver, we systematically explore the parameter space through 600+ equilibrium calculations spanning triangularity ($-0.7 \leq \delta \leq 0.7$) and elongation ($1.0 \leq \kappa \leq 2.1$) while respecting SPARC's coil current limits and first-wall constraints and selecting a reduced-field 8~T operating point. We find that moderate NT double-null plasmas ($\delta = -0.35$, $\kappa = 1.68$) are achievable at 8~T with plasma currents of $\sim$2.1 MA. However, NT operation requires 43\% plasma volume reduction (from $\sim$20.0 to $\sim$11.4 m$^3$) compared to the baseline PT design due to vessel walls optimized for PT conformity, and connection lengths are reduced by 40\% due to geometric mismatch with the PT-optimized divertor. Central solenoid current requirements decrease by 54\% in NT configurations, though specific shaping coils (PF$_3$) require a factor of $\sim$5.5 higher currents. All equilibria satisfy fundamental MHD stability criteria with comfortable margins. Thus, while performance is degraded relative to baseline design, these results demonstrate SPARC's potential as a high-field experimental bridge between contemporary NT experiments and proposed NT reactor concepts, capable of testing whether NT's operational advantages (ELM-free operation, favorable impurity transport) persist under reactor-relevant conditions provided by SPARC.\footnote{Corresponding author: Narin Y\"uksek, narin015@mit.edu.  Jan. 10, 2026}
\end{abstract}

\section{Introduction}
Negative triangularity (NT) tokamak configurations have re-emerged as a promising pathway toward reactor-relevant plasmas with improved power handling and reduced edge turbulence, without reliance on H-mode operation \cite{Marinoni_2021_review}. Experimental studies on TCV \cite{Camenen_2007_TCV, Fontana_2017_TCV, Huang_2018_TCV, Faitsch_2018_TCV, Coda_2021_TCV, Han_2021_TCV, Balestri_2024_TCV, Riencker_2025_TCV, Morgan_2025_TCV}, DIII-D \cite{Austin_2019_DIIID, Marinoni_2019_DIIID, Saarelma_2021_DIIID, PazSoldan_2024_DIIID, Thome_2024_DIIID, Casali_2025_DIIID}, and ASDEX Upgrade (AUG) \cite{Happel_2022_AUG} have demonstrated enhanced confinement and ELM-free operation in sufficiently negative triangularity plasmas, motivating renewed interest in NT for future fusion devices. In parallel, conceptual reactor studies have proposed NT as an attractive design option for high-field reactors, 
and several reactor-scale NT concepts have been proposed in recent years. Kikuchi \emph{et al.} \cite{Kikuchi_2019_reactor} explored NT configurations for L-mode steady-state tokamak reactors, demonstrating potential advantages in current drive efficiency. De Boucaud \emph{et al.} \cite{deBoucaud_2024_ARC} explored the NT equilibria on the ARC device \cite{Kuang_2018_ARC}, while the MANTA concept \cite{Rutherford_2024_MANTA} represents a purpose-built NT reactor design in the ARC family. More general studies have further explored the design space and developed tools for NT reactors, \emph{e.g.} \cite{Medvedev_2015_reactor, Sauter_2016_reactor, Wilson_2025_reactor}.

In this study, we explore SPARC: a compact ($R_0 = 1.85 m$), high-field tokamak currently under construction, designed to demonstrate net fusion energy with positive triangularity plasmas \cite{RodriguezFernandez_2022_SPARC}. The 12.2~T on-axis toroidal field of the baseline SPARC V2 design is made possible by the development of REBCO high-temperature superconducting magnets \cite{Hartwig_2024_HTS, Sanabria_2024_HTS}, while the REBCO central solenoid will drive a nominal plasma current of $I_p= 8.7$~MA \cite{Creely_2020_SPARC, RodriguezFernandez_2022_SPARC}. The poloidal field (PF) coil system comprises four coil pairs, symmetric about the midplane, while the central solenoid (CS) has four pairs of coils.  Copper divertor coils are situated near the plasma, just outside the vacuum vessel.  The magnets and vacuum vessel are optimized for the nominal PT double-null configurations. The vessel features tight-fitting first walls ($R_{min} = 1.27$~m to $R_{max} = 2.43$~m at the midplane) that constrain achievable plasma shapes. 

While negative triangularity concepts have appeared for ARC qualitatively in recent community-facing presentations and conference posters (\emph{e.g.}, SOFE 2025), the possibility of SPARC demonstrating such configurations is new, as far as the authors are aware.  Exploring NT operation in SPARC could provide a bridge between experiments on contemporary devices including TCV, DIII-D, and AUG and the proposed reactor concepts cited above, validating this operational mode in a device with reactor-relevant engineering constraints.

However, this raises a key question: can the SPARC tokamak, with its tight geometric constraints, support NT equilibria, and within what limits? 


In this paper, we present a systematic computational study of NT equilibria in SPARC geometry using the FreeGS free-boundary Grad-Shafranov solver\cite{FreeGS}.  Section 2 describes our computational methods, including plasma parameterization. Section 3 presents results from parameter space scans spanning triangularity ($\delta$) of [-0.5, 0.5] and elongation ($\kappa$) of [1.0, 2.0], identifying achievable configurations and limiting factors. Section 4 discusses these results, and Section 5 summarizes our key findings. 


\section{Methods}
We compute axisymmetric MHD equilibria using FreeGS \cite{FreeGS}, an open-source Python library for solving the Grad–Shafranov equation \ref{eq:grad_shafranov} with free boundaries \cite{Freidberg_2014_book}. 

The solver iterates between the plasma current distribution and the magnetic flux from external coils until a self-consistent equilibrium is achieved where the plasma boundary satisfies force balance conditions.

\begin{equation} \label{eq:force_balance}
\nabla p=\mathbf{J}\times\mathbf{B}
\end{equation}

Equation \ref{eq:force_balance} gives the force balance that must be satisfied for equilibrium.  In the axisymmetric geometry, this gives rise to the Grad-Shafranov equation, \ref{eq:grad_shafranov}

\begin{equation} \label{eq:grad_shafranov}
\frac{\partial^2 \psi}{\partial R^2}-\frac{1}{R}\frac{\partial \psi}{\partial R} + \frac{\partial^2 \psi}{\partial z^2}=-\mu_0 R^2 \frac{dp}{d\psi}-\frac{1}{2} \frac{dF^2}{d\psi}
\end{equation}
where $p(\psi)$ is the pressure, given as a flux function; $F(\psi)\equiv R B_{\phi}$, and
\begin{equation}
\mathbf{B}=\frac{1}{R}\nabla\psi\times\hat{\mathbf{e}}_{\phi}+\frac{F}{R}\hat{\mathbf{e}}_{\phi}
\end{equation}
and
\begin{equation}
\mu_0 \mathbf{J}=\frac{1}{R} \frac{dF}{d\psi} \nabla\psi\times\hat{\mathbf{e}}_{\phi}-\left[\frac{\partial}{\partial R}\left(\frac{1}{R} \frac{\partial\phi}{\partial R}\right)+\frac{1}{R} \frac{\partial^2\psi}{\partial z^2}\right]\hat{\mathbf{e}}_{\phi}
\end{equation}

Pixel interpolation from the device footprint published in \cite{Creely_2020_SPARC, RodriguezFernandez_2022_SPARC} was used to determine the coordinates of the wall and coils for SPARC V2.

For a given set of shaping parameters, (triangularity, $\delta$; elongation, $\kappa$; geometric centroid, $R_0$, $z_0$; minor radius, $r$) the desired locations of the x-points and the outer and inner midplane points are calculated using the Miller flux surface parameterization, \cite{Miller_1998}, as shown in Eq. \ref{eq:Miller},

\begin{equation} \label{eq:Miller}
    \begin{split}
    R_s&=R_0+r \cos{\left[\theta+\left(\sin^{-1}\delta\right)\sin\theta\right]} \\       
    z_s&=\kappa r \sin \theta
    \end{split}
\end{equation}
subject to the size constraints of the device,
\begin{equation}
    \begin{split}
        R_s(\theta=0)&= R_0+r < R_{max} = 2.43\,\mathrm{m} \\
        R_s(\theta=\pi)&= R_0-r > R_{min} = 1.27\,\mathrm{m} \\
        r&=\frac{z_{max}}{\kappa}< \frac{R_{max}-R_{min}}{2} = 0.58\,\mathrm{m} \\
    \end{split}
\end{equation}
For the nominal discharge, $R_0=\frac{R_{max}+R_{min}}{2} = 1.85\,\mathrm{m}$.

Four points on the last-closed flux surface (LCFS) were specified: inner midplane, outer midplane, upper x-point, and lower x-point.  In some cases, two to four additional points on the LCFS were specified at various poloidal angles.  Additional shaping constraints were explored including strike point specifications at both inner and outer divertor targets. However, including strike points alongside x-point and isoflux constraints often failed to converge with reasonable coil currents.  A minimal set of constraints provided a more robust convergence, sacrificing some fine control of the plasma shape.

Lastly, current limits were applied on all poloidal field coils, divertors, and central solenoid circuits.

Given the fact that SPARC is not optimized for NT geometry, with the divertor of minimal use for these plasmas, a low performance plasma was sought to relieve wall loading.  As such, we investigated reduced-field scenarios with $B_0$ = 8~T and $R_0$ = 1.75~m.

To enable more direct comparison between NT and PT performance, we computed 8~T PT equilibria with identical field strength, major radius, and current scaling as the NT cases. Comparing NT directly against SPARC's 12.2~T baseline would conflate shaping effects with differences in field strength and current. The matched 8~T PT cases isolate the impact of triangularity sign, allowing us to attribute differences in achievable parameter space and coil requirements specifically to the change in $\delta$ rather than to variations in the magnetic field configuration.

The specific elongation and triangularity values presented were determined through systematic parameter scans spanning 600+ test cases across $\delta \in [-0.7, 0.7]$ and $\kappa \in [1.0, 2.1]$ with many configurations failing to converge.

\subsection{Scaling plasma current and pressure}
As per Creely \emph{et al.} \cite{Creely_2020_SPARC}, we utilize the Uckan estimate for $q$,

\begin{equation}
q_{Uckan}^*=\frac{5a^2 B_0}{R_0 I_p}\frac{1+\kappa_{95}^2\left(1+2\delta_{95}^2-1.2\delta_{95}^3\right)}{2}
\end{equation}
to constrain the plasma current
\begin{equation} \label{eq:Ip}
\boxed{
I_p=\frac{5a^2 B_0}{R_0 q_{Uckan}^*}\frac{1+\kappa_{95}^2\left(1+2\delta_{95}^2-1.2\delta_{95}^3\right)}{2}}
\end{equation}
where $I_p$ is in MA.  $q_{Uckan}$ is then an input to each computation.

The core pressure was determined from simplified profiles given in \cite{Freidberg_2015}
\begin{equation} \label{eq:pressure_profile}
p=\bar{p}\left(1+\nu_p\right)\left(1-\rho^2\right)^{\nu_p}=2.5\bar{p}\left(1-\rho^2\right)^{3/2}
\end{equation}
for which the core value may be written in terms of the average pressure as
\begin{equation}\label{eq:avg_to_core}
p_0=2.5\bar{p}
\end{equation}

We then write $\beta$ in terms of the average pressure and $B_0$, as follows
\begin{equation}
\beta=\frac{4.02\times10^{-3}\bar{p}}{B_0^2}
\end{equation}
and we solve for the average pressure:
\begin{equation}
\bar{p}=\beta\frac{B_0^2}{4.02\times10^{-3}}
\end{equation}
from which Eq. \ref{eq:avg_to_core} gives the core pressure value
\begin{equation}
\boxed{
p_0=595\beta B_0^2}
\end{equation}

Here, pressure is expressed in units of keV$\cdot 10^{19}$~m$^{-3}$.  Expressing $p$ in units of Pa,

\begin{equation}
\boxed{
p_0\approx 10^6 \beta B_\phi^2}
\end{equation}




For all cases, a $\beta$ set point of $0.007$ was used; this corresponds to the 12~T L-Mode case from Creely \emph{et al.} \cite{Creely_2020_SPARC}.  For the NT case, the set point for $q_{Uckan}$ was $5.5$; this value was chosen to be higher than the PRD value of $3.05$ for a lower plasma current than the PRD. These selections were made again in order to reduce plasma performance given that power handling will not be optimized for NT in SPARC. They also improve equilibrium convergence within coil constraints.

\subsection{Stability Limits}

We assess whether our equilibria satisfy fundamental MHD stability constraints: the Troyon beta limit and the kink safety factor limit \cite{Freidberg_2015}. 

The maximum achievable $\beta$ given by the Troyon limit:

\begin{equation}
\beta < \beta_T = \beta_N \frac{I_p}{aB_0}, \quad \beta_N = 2.8\%
\end{equation}

where $I_p$ is in MA, $a$ is the minor radius in m, and $B_0$ is the toroidal field in T.

\textbf{For the PRD 12.2 T PT case:}
\begin{itemize}
    \item $I_p = 8.7$ MA
    \item $B_0 = 12.2$ T
    \item $a = 0.57$ m
    \item $\beta = 0.00368$
\end{itemize}

\begin{equation}
\beta_T = 0.028 \times \frac{8.7}{0.57 \times 12.2} = 0.035
\end{equation}

$0.00368 < 0.035$ (Satisfied with substantial margin)

\textbf{For the PT 8 T case:}
\begin{itemize}
    \item $I_p = 1.99$ MA
    \item $B_0 = 8$ T
    \item $a = 0.45$ m
    \item $\beta = 0.00216$
\end{itemize}

\begin{equation}
\beta_T = 0.028 \times \frac{1.99}{0.45 \times 8} = 0.0155
\end{equation}

$0.00216 < 0.0155$ (Satisfied with substantial margin)

\textbf{For the NT 8 T case:}
\begin{itemize}
    \item $I_p = 2.1$ MA
    \item $B_0 = 8$ T
    \item $a = 0.45$ m
    \item $\beta = 0.00218$
\end{itemize}

\begin{equation}
\beta_T = 0.028 \times \frac{2.1}{0.45 \times 8} = 0.0163
\end{equation}

Check: $0.00218 < 0.0163$ (Satisfied with substantial margin)

To avoid current-driven major disruptions, the edge safety factor must satisfy the kink safety factor limit:

\begin{equation}
q_* > q_K = 2
\end{equation}

All three cases satisfy this criterion with their computed edge safety factors: $q_{95} = 2.83$ for the PRD case, $q_{95} = 4$ for the PT 8 T case, and $q_{95} = 3.9$ for the NT 8 T case. 

Combined with the Troyon beta limit analysis, all three equilibria demonstrate comfortable stability margins, with the reduced current in the 8 T cases providing even greater margin against beta-limited disruptions.

\section{Results}
\begin{figure}
\centering
\includegraphics[width=0.5\textwidth]{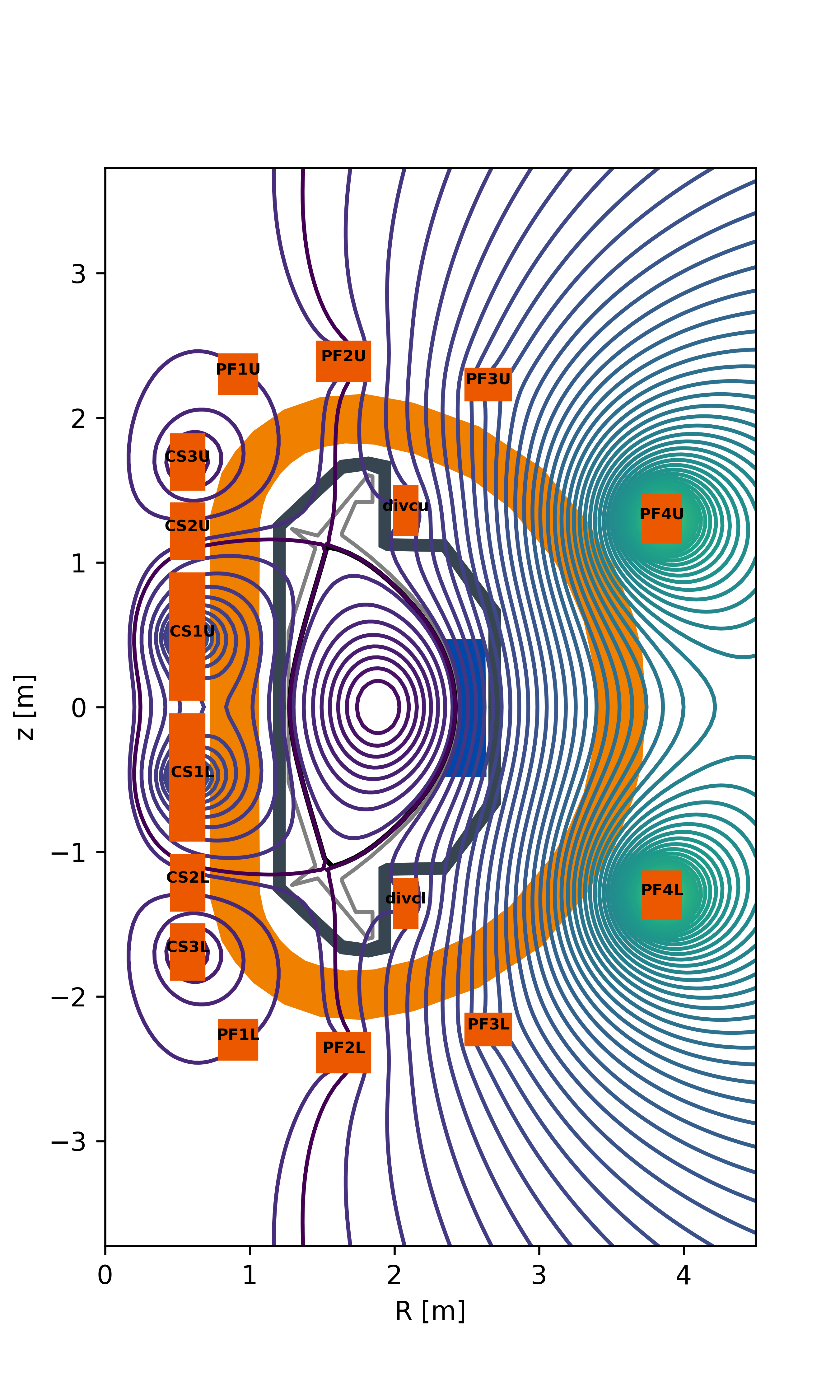}
\caption{FreeGS equilibrium for the nominal 12.2~T SPARC V2 PRD case overlaid with SPARC layout with PF, CS, and divertor coils labeled.}
\label{fig:Nom_equil}
\end{figure}

\begin{figure}
\centering
\includegraphics[width=0.5\textwidth]{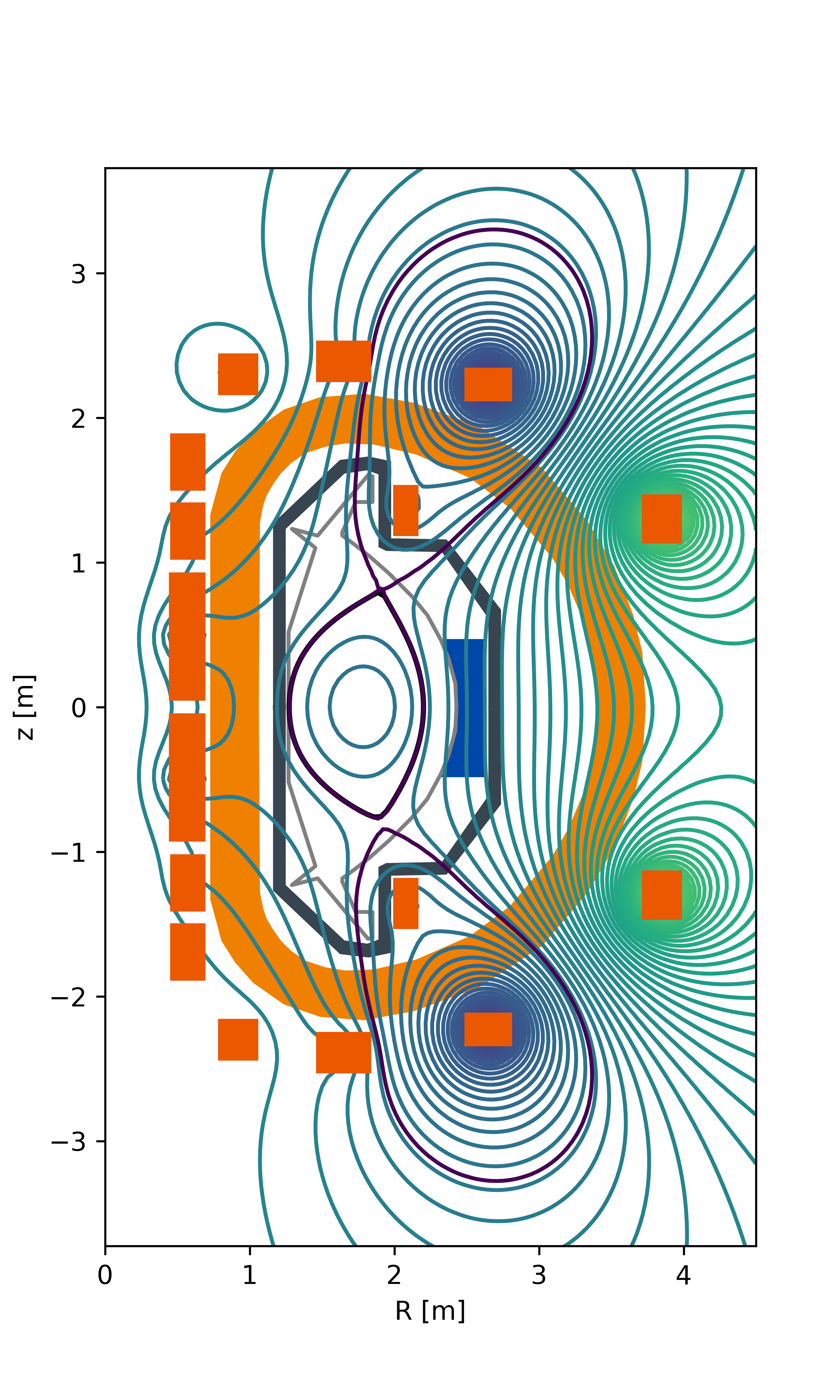}
\caption{FreeGS equilibrium for the NT 8~T case with $\delta=-0.37$ and $\kappa=1.68$.}
\label{fig:NT 8T}
\end{figure}

\begin{figure}
\centering
\includegraphics[width=0.5\textwidth]{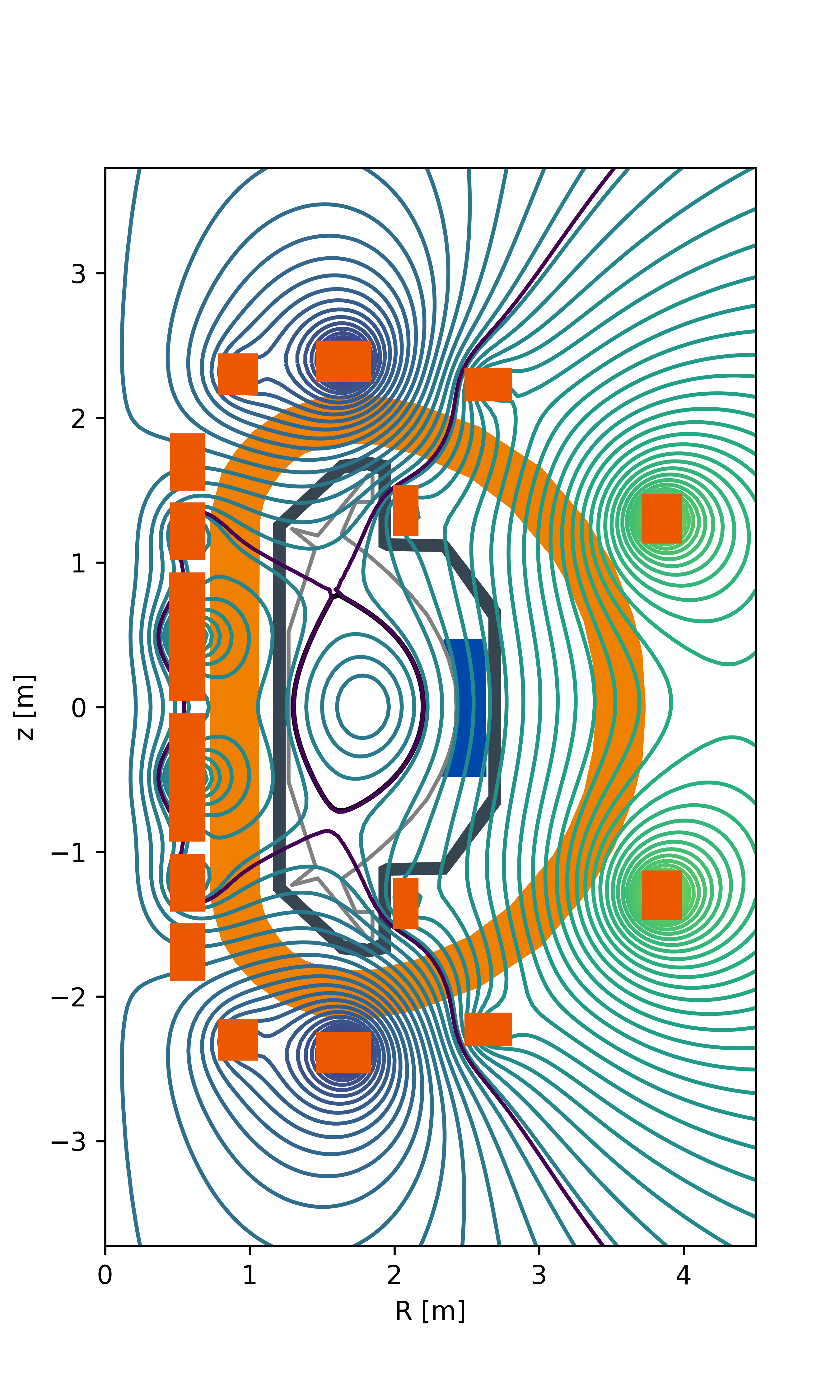}
\caption{FreeGS equilibrium for the PT 8~T case with $\delta=3.7$ and $\kappa=1.75$.}
\label{fig:PT 8T}
\end{figure}

 Three cases are presented below showing configurations from the parameter scan that achieved convergence with coil currents within engineering limits and adequate plasma-wall clearance.  They are: 
\begin{enumerate}
    \item The nominal 12.2~T PT case (PRD)  (Fig. \ref{fig:Nom_equil})
    \item An NT 8~T case (Fig. \ref{fig:NT 8T})
    \item A PT 8~T case (Fig. \ref{fig:PT 8T})
\end{enumerate}
 
\begin{figure}
\centering
\includegraphics[width=0.5\textwidth]{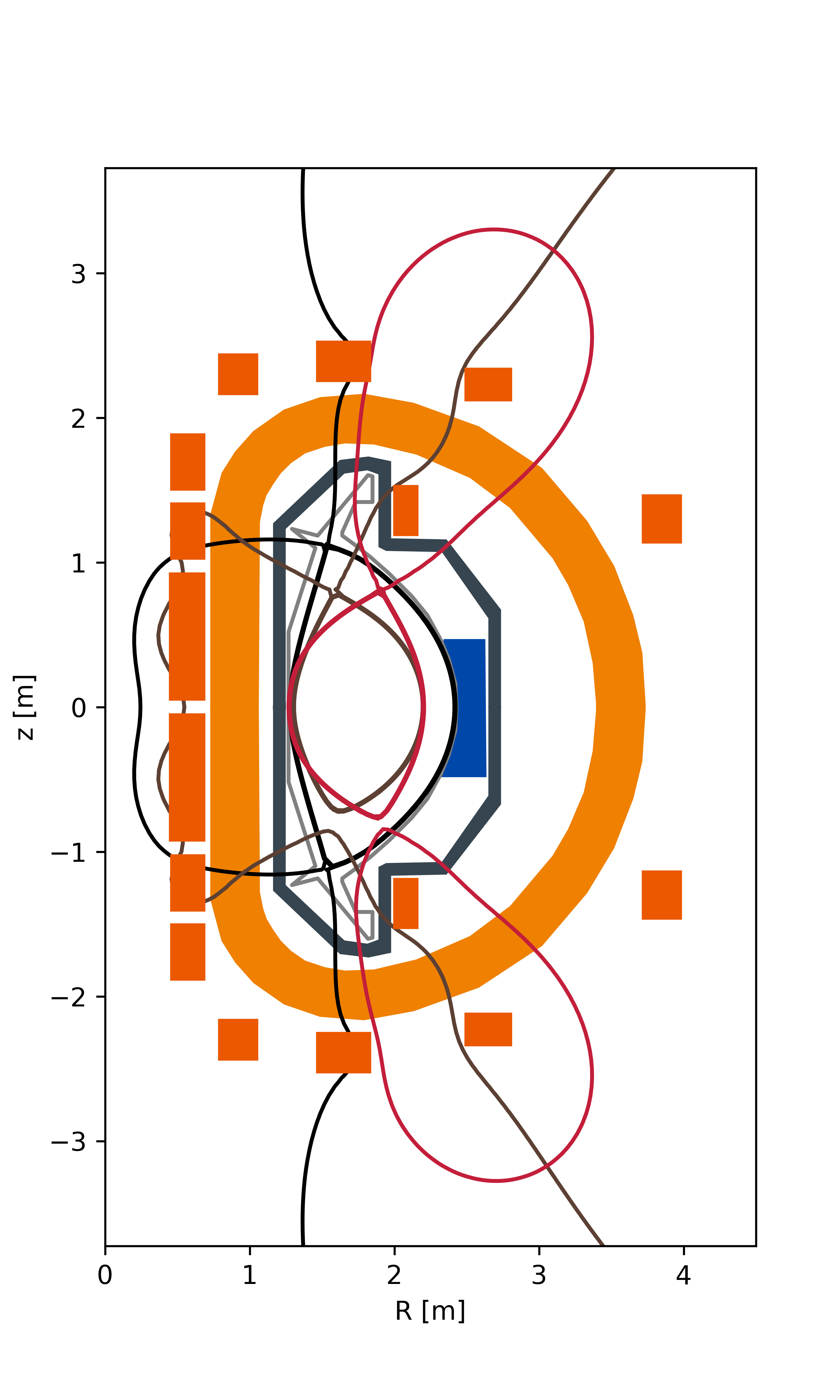}
\caption{Separatrix for all three cases: the black line corresponds to the PRD case, the brown line corresponds to the 8~T PT case, and the red line corresponds to the 8~T NT case.}
\label{fig:all_LCFS}
\end{figure}

 Parameters from these cases are summarized in Table \ref{tab:case_par}.  And Figure~\ref{fig:all_LCFS} plots the LCFS from all three cases. More generally, we found that elongations in the range $\kappa \in [1.65, 2.0]$ and triangularities in the range $|\delta| \in [0.3, 0.6]$ generally converged most reliably for both NT and PT scenarios, though the accessible parameter space differed between the two due to their distinct magnetic field topologies and associated control requirements.

Before exploring the NT parameter space, we validated our \texttt{FreeGS} implementation against the baseline PT case. Figure \ref{fig:Nom_equil} shows a reproduction of the nominal 12.2~T PRD equilibrium that approximates the case described in \cite{Creely_2020_SPARC} and \cite{RodriguezFernandez_2022_SPARC}. 

We then produced double-null equilibria of reduced-field (8~T) PT and NT cases. Global plasma parameters and coil currents are shown in Table \ref{tab:case_par}.

A minimal set of success criteria we looked for are: 
\begin{enumerate}
    \item \texttt{FreeGS} converges upon an equilibrium solution
    \item All coil currents for that equilibrium are within engineering limits
    \item X-points and shaping parameters are close to target values
    \item The plasma boundary remains within first wall.
\end{enumerate}

To accommodate NT within a vessel optimized for PT, the plasma size had to be reduced in order to keep the plasma, and particularly the X-points, inside the vessel walls with non-vanishing divertor leg lengths. As such, the major and minor radii and elongation were reduced from PRD values of 1.85~m, 0.57~m, and $\sim2$ to 1.75~m, 0.4-0.45~m, and 1.63-1.75 in 8~T NT scenarios. These NT values were then duplicated in a PT 8 T case for comparison. This also resulted in a volume reduction of 47\% in the featured PT case (10.30~m$^3$) and a 42\% reduction in NT (11.37~m$^3$).

However, while the vessel footprint necessitated a reduction in size for NT, this tended to increase demand from PF coil currents.  The shaping parameters here seek a balance between these design constraints. Indeed, higher elongation improved current margins by increasing plasma volume and reducing the demands on shaping coils. However, elongating the plasma excessively led to plasma–wall contact, X-points approaching the wall, or transitions to single-null configurations in cases where double-null or near-double-null operation was desired. Conversely, excessive reduction in plasma dimensions, while maintaining safe standoff from material surfaces, tended to inflate current requirements beyond acceptable values.

The total number of amp-turns in the poloidal field coils for the PT 8~T case was 19.16 MA-turns, compared with 22.34 MA-turns in NT 8~T. However, the total number of amp-turns in the central solenoid coils for the PT 8~T case was 26 MA-turns, while it was 11.98 MA-turns in NT 8T. For the PRD, the total number of amp-turns in the PF coils was 21.66 MA-turns, and for the CS coils, 51.3 MA-turns. We see that while there are roughly comparable total currents in the PF coils in all cases, the currents in the CS coils are substantially higher in the PT cases. Meanwhile, PF3 is consistently much higher in the NT cases. This discrepancy is not unexpected given the shift of x-points from the high-field side in the PT case to the low-field side in the NT case.

Connection length provides insight into power handling characteristics in the scrape-off layer. We computed connection lengths by field line tracing starting 1 mm outside the last closed flux surface at the outboard midplane. 
The NT 8 T case exhibits a connection length of 17.5~m, significantly shorter than the PT cases (27.5~m for PRD and 29.4~m for PT 8 T), reflecting the geometric mismatch. This constraint, combined with coil current limitations, motivated deliberate operation at lower $\beta$ in the NT scenarios. 
While shorter divertor legs are less favorable for detachment, NT may compensate through intrinsically lower edge and divertor temperatures arising from a lack of an edge pedestal, and this may facilitate detachment.

We assess whether our equilibria satisfy fundamental MHD stability constraints: the Troyon beta limit and the kink safety factor limit \cite{Freidberg_2015}.  The Troyon $\beta$ limit,

\begin{equation}
\beta < \beta_T = \beta_N \frac{I_p}{aB_0}, \quad \beta_N = 2.8\%
\end{equation}
is well-satisfied in all cases, while the kink safety factor is adequately high.

\begin{table*}
\centering
\caption{Comparison of Equilibrium Parameters}
\label{tab:case_par}
\begin{tabular}{|l|l|l|l|}
\hline
\textbf{Parameter} & \textbf{PRD (12.2 T, PT)} & \textbf{PT 8 T} & \textbf{NT 8 T} \\\hline
\textbf{$\delta$} & 0.52 & 0.35 & -0.35 \\\hline
\textbf{$\kappa$} & 1.94 & 1.68 & 1.68 \\\hline
\textbf{$R_0$}~[m] & 1.85 & 1.75 & 1.75 \\\hline
\textbf{$a$}~[m] & 0.57 & 0.45 & 0.45 \\\hline
\textbf{$I_p$}~[MA] & -8.7 & -1.99 & -2.1 \\\hline
\textbf{$B_0$}~[T] & 0 & 0 & 0 \\\hline
\textbf{$V$}~[m$^3$] & 20.0 & 10.3 & 11.4 \\\hline
\textbf{$\beta$} & 0.00368 & 0.00216 & 0.00218  \\\hline
\textbf{$\beta_N$} & 0.29\% & 0.39\% & 0.36\% \\\hline
\textbf{$q_{95}$} & 2.83 & 4 & 3.9 \\\hline
\textbf{Connection length}~[m] & 27.5 & 29.4 & 17.5 \\\hline
\textbf{PF$_1$}~[MA-turns] (U/L) & 0.09 & -2.5 / 2.5 & 2 / 0.8  \\\hline
\textbf{PF$_2$}~[MA-turns] (U/L) & -0.58 & -4.3 / -4.3 & -4.3 / 0.94  \\\hline
\textbf{PF$_3$}~[MA-turns] (U/L) & 0.31 & 0.96 / 1 & -5.3 / -5.3  \\\hline
\textbf{PF$_4$}~[MA-turns] (U/L) & 5.9 & 1.8 / 1.9 & 3.9 / 3.8 \\\hline
\textbf{divc}~[MA-turns] (U/L) & -0.75 & 0.75 / 0.72 & -0.75 / -0.75 \\\hline
\textbf{CS$_1$}~[MA-turns] (Inner/Outer) & 4.8 / 13.13 & 2.6 / 5.8 & -2.9 / -2.4  \\\hline
\textbf{CS$_2$}~[MA-turns] & 0 & 4.6 & -0.69  \\\hline
\textbf{CS$_3$}~[MA-turns] & -7.89 & 0 & 0 \\\hline
\end{tabular}
\end{table*}

\section{Discussion}

The geometric constraints imposed by SPARC's PT-optimized design necessitate significant trade-offs for NT operation. The 47\% plasma volume reduction directly impacts fusion performance. Combined with the deliberately conservative $\beta$ = 0.007 setting chosen to reduce power handling demands and improve equilibrium convergence, these NT scenarios would produce substantially lower fusion power than SPARC's primary mission parameters in DT operation.

However, NT operation reveals potentially advantageous characteristics in the coil current distribution. The central solenoid requirements in the NT 8~T case (11.98 MA-turns total) are reduced by 54\% compared to the PT 8~T case (26 MA-turns) and by 77\% compared to the PRD case (51.3~MA-turns). This dramatic reduction in CS current demand could translate to extended pulse capability, reduced power supply requirements, or more compact central solenoid designs in purpose-built NT reactors. This advantage must be weighed against the substantially increased demands on specific shaping coils, particularly $PF_3$, which requires -5.3 MA-turns in NT compared to 0.96 MA-turns in PT 8~T--a factor of ~5.5 increase. The total poloidal field coil current (excluding CS) is comparable across cases: 21.66 MA-turns for PRD, 19.16 MA-turns for PT 8 T, and 22.34 MA-turns for NT 8~T. However, the distribution of these currents differs, reflecting the distinct magnetic topology requirements of NT configurations, and highlighting the need to optimize the coil design and disposition to a particular baseline.


\section{Conclusion}

This study explored the feasibility of producing NT equilibria in SPARC. The central question we addressed was whether a device with a tight-fitting first wall, PT-optimized divertor and poloidal field coil system could accommodate NT within its engineering constraints.

This presented significant challenges. SPARC's vessel geometry features tight radial limits ($R_{min} = 1.27$~m to $R_{max} = 2.43$ m at the midplane). The divertor targets and x-point locations are positioned for high-field-side x-points characteristic of PT, while NT naturally produces low-field-side x-points. The coil locations were also optimized for a PT baseline. These constraints meant that achieving NT equilibria required navigating a narrow parameter space where plasma shaping demands, coil current limits, and plasma-wall clearance requirements could all be simultaneously satisfied.

By reducing the toroidal field to 8~T, decreasing the major radius from 1.85~m to 1.75~m, reducing the minor radius to 0.4~ m, lowering the elongation to $\kappa$ = 1.68, and operating at conservative plasma current (2.1~MA) and $\beta$ (0.007), we successfully produced NT equilibria within the SPARC footprint while respecting all coil current limits and maintaining adequate wall clearance. Through parameter space exploration spanning 600+ equilibrium calculations across triangularity $\delta \in [-0.7, 0.7]$ and $\kappa \in [1.0, 2.1]$, we identified the accessible NT operating regime and characterized its properties.

Comparing the three primary cases analyzed:

\begin{itemize}
    \item \textbf{PRD (12.2~T, PT):} The baseline design is characterized by $I_p$=8.7 MA, $\delta$=0.52, $\kappa$=1.94, with plasma volume $V$=19.97~m${^3}$, representing SPARC's target high-performance operating point with connection length 27.5~m and requiring 51.3~MA-turns total in the central solenoid.
    \item \textbf{PT 8 T:} The reduced-field $+\delta$ comparison case operates at $I_p$=1.99 MA, $\delta$ = 0.35, $\kappa$=1.68, with $V$=10.30~m${^3}$ (47\% volume reduction), connection length 29.4~m, and requires 26~MA-turns in the central solenoid with relatively modest PF coil currents.
    \item \textbf{NT 8 T:} The $-\delta$ case achieves $I_p$=2.1~MA, $\delta$=-0.35, $\kappa$=1.68, with $V$=11.37~m${^3}$ (42\% volume reduction), connection length 17.5~m, and requires only 11.98~MA-turns in the central solenoid but dramatically increased $PF_3$ current (-5.3 MA-turns, a factor of $\sim$5.5 higher than PT cases).
\end{itemize}

All three equilibria satisfy fundamental MHD stability criteria with comfortable margins, operating well below Troyon beta limits and above kink safety factor thresholds.

These results demonstrate that while SPARC was not designed for NT operation and requires significant operational modifications to accommodate it, NT equilibria remain feasible within the device's engineering envelope.

SPARC could provide crucial validation of whether the intrinsic physics benefits of NT -- the absence of an edge pedestal leading to naturally lower edge temperatures, the enhanced outward impurity transport, the elimination of ELM-driven transient heat loads, the lack of an LH power threshold, and so on -- are manifested in such a high-performance device.

Moreover, SPARC would serve as an essential experimental bridge connecting contemporary experiments that have established NT's fundamental advantages at modest scale and field, and proposed high-field NT reactor concepts that rely on these advantages scaling favorably to reactor conditions.  

For future reactor design, our analysis underscores the value of purpose-built NT optimization. Dedicated NT reactors could avoid the volume reductions and divertor geometry compromises we encountered by designing vessel shapes, first-wall contours, and poloidal field coil systems specifically for NT magnetic topology. The reduced central solenoid requirements  and the potential for enhanced impurity transport suggest genuine advantages for compact reactor concepts, while the need for enhanced shaping coil capacity and optimized divertor configurations identifies key design considerations that purpose-built devices should address from the outset.

Immediate next steps for future work may include: 
\begin{enumerate}
    \item Upper and lower single-null PT/NT cases for alternative divertor configurations that might better accommodate NT
    \item \texttt{BALOO} edge turbulence simulations using related equilibria
    \item \texttt{FreeGSNKE} kinetic equilibrium calculations
    \item Vertical stability analysis using \texttt{MEQ} (\texttt{MATLAB} equilibrium toolbox) to assess controllability of high-$\kappa$ NT configurations
    \item Integrated scenario modeling with \texttt{OMFIT} to simulate proposed experimental discharges 
\end{enumerate}

\section{Acknowledgments}
The authors extend their appreciation to Lauren Milechin, Juliana Mullen, Emmanuel Lanti, Mark London, and Lee Berkowitz for their support with the technical and organizational aspects of this project, as well as their patience and encouragement in navigating countless emails and logistical details that ensured the practical foundations of this work were solid and reliable.

We also thank Kathreen Thome for sharing her presentation on NT in APS DPP 2025.

Lastly, we gratefully acknowledge the support of Lauren Bandklayder and Christian Theiler for their steady presence and willingness to facilitate opportunities that played an important role in shaping the path that led to this work.

\bibliography{bibliography.bib}
\bibliographystyle{unsrturl}

\end{document}